\documentclass[a4paper,11pt]{article}
\usepackage{pos}
\usepackage[capitalize]{cleveref}
\usepackage{siunitx}
\usepackage{xcolor}
\usepackage{amsmath}

\title{Status and Developments in Polarised Parton Distribution Functions}
\ShortTitle{Developments in Polarised Parton Distribution Functions}

\author*[a]{Amedeo Chiefa}

\affiliation[a]{
  \it The Higgs Centre for Theoretical Physics, University of Edinburgh,\\
      JCMB, KB, Mayfield Rd, Edinburgh EH9 3JZ, Scotland}

\emailAdd{amedeo.chiefa@ed.ac.uk}

\abstract{ The need for accurate and precise polarised parton distribution
          functions (PDFs) is becoming increasingly crucial in view of the
          Electron-Ion Collider experimental program foreseen in the coming
          years. Two global PDF determinations at next-to-next-to-leading order
          accuracy have been recently presented, {\sc MAPPDFpol1.0} and BDSSV24.
          I review the former and I provide a comparative discussion of other
          PDF sets accurate to next-to-leading order. I show that differences
          between these PDF sets, due to the choice of experimental and
          methodological input, exceed differences due to perturbative accuracy.
          }

\FullConference{31st International Workshop on Deep Inelastic Scattering (DIS2024)\\
 8–12 April 2024\\
Grenoble, France\\}

\begin{document}
\maketitle

\section{Introduction}

Helicity-dependent parton distribution functions (PDFs)~\cite{Ethier:2020way}
are essential for understanding the nucleon spin structure in terms of quarks
and gluons~\cite{Ji:2020ena}. Following the pioneering experiments on polarised
deep inelastic scattering (DIS) conducted by the European Muon
Collaboration~\cite{EuropeanMuon:1987isl,EuropeanMuon:1989yki}, efforts in this
field have advanced significantly. The upcoming Electron-Ion Collider
(EIC)~\cite{AbdulKhalek:2021gbh,AbdulKhalek:2022hcn}, expected to operate in the
2030s, promises substantial improvements in measuring observables for
longitudinally polarised inclusive and semi-inclusive DIS (SIDIS), extending the
probed kinematic region (see {\it e.g.}, Fig.~1 in Ref.~\cite{Ethier:2020way})
and achieving higher measurement precision.

Future EIC measurements, anticipated to reach the percent-level precision
~\cite{Borsa:2020lsz}, require equally accurate theoretical predictions and
PDFs. While there have been several recent advancements in perturbative
calculations for both DIS~\cite{Blumlein:2022gpp,
Hekhorn:2018ywm,Behring:2015zaa,Ablinger:2019etw,Behring:2021asx,Blumlein:2021xlc,
Bierenbaum:2022biv,Ablinger:2023ahe} and
SIDIS~\cite{Abele:2021nyo,Bonino:2024wgg, Goyal:2024tmo} structure functions
\footnote{In addition, very recent calculations for three-loop matrix elements in the
  polarised case using the Larin-scheme~\cite{Blumlein:2024euz} have been
  presented in Ref.~\cite{Ablinger:2024xtt}.}, the bulk of the available polarised
PDF sets remains, to date, at next-to-leading order
(NLO)~\cite{deFlorian:2014yva, Nocera:2014gqa,Ethier:2017zbq}.

Very recently, two new NNLO analyses of polarised PDFs have been presented in
Refs.~\cite{Borsa:2024mss,Bertone:2024taw}, although only the PDF sets in
Ref.~\cite{Bertone:2024taw} have been made public. Both determinations are based
on available DIS and SIDIS data, but Ref.~\cite{Borsa:2024mss} also includes
proton-proton ($pp$) spin asymmetry data from the Relativistic Heavy Ion
Collider (RHIC). These determinations extend a previous NNLO analysis based on
solely DIS data~\cite{Taghavi-Shahri:2016idw}, and will be followed by a new
NNLO determination~\cite{Hekhorn:2024jrj}. Here, I briefly discuss {\sc
MAPPDFpol1.0}, the new polarised PDF set presented in
Ref.~\cite{Bertone:2024taw}, before moving to a comparative analysis of the
currently available polarised PDF sets at NLO.

\section{{\sc \bf{MAPPDFpol1.0}} at NNLO}
\label{sec:NNLO}

The polarised PDF set {\sc MAPPDFpol1.0} presented in
Ref.~\cite{Bertone:2024taw} leverages a fitting framework that combines a
neural-network parametrisation~ \cite{AbdulKhalek:2020uza} of polarised PDFs
with a Monte Carlo representation of PDF uncertainties, aiming at reducing
parametrisation bias as much as possible and at obtaining statistically sound
uncertainties.

The experimental information included in {\sc MAPPDFpol1.0} relies on the
currently available structure function data for
inclusive~\cite{EuropeanMuon:1989yki,SpinMuon:1998eqa,
COMPASS:2015mhb,COMPASS:2016jwv,E142:1996thl,E143:1998hbs,E154:1997xfa,E155:2000qdr,
HERMES:1997hjr,HERMES:2006jyl,JeffersonLabHallA:2016neg,CLAS:2014qtg} and
semi-inclusive~\cite{COMPASS:2010hwr,HERMES:2018awh} DIS. In addition, SU(2) and
SU(3) flavour symmetries are assumed to hold exactly, so that the lowest moments
of the triplet and octet polarised PDF combinations can be compared to the
values extracted from semi-leptonic
$\beta$-decays~\cite{ParticleDataGroup:2022pth}. The parametrised flavours are
$\Delta f_u$, $\Delta f_{\bar{u}}$, $\Delta f_d$, $\Delta f_{\bar{d}}$, $\Delta
f_s$, $\Delta f_{\bar{s}}$, and $\Delta f_g$. The initial parametrisation scale
is $Q^2=1$ GeV$^2$. Note that $\Delta f_s$ and $\Delta f_{\bar{s}}$ are
parametrised separately thanks to the charged kaon production in SIDIS data and
to the NNLO corrections which make strange quark and antiquark evolve
differently. Finally, the positivity constraint of cross-sections is enforced
through the PDFs by imposing $|\Delta f| \leq f$.

Theoretical predictions are computed using the public code {\sc
APFEL}++~\cite{Bertone:2013vaa,Bertone:2017gds}. Polarised coefficient functions
for DIS are implemented using exact calculations up to
NNLO~\cite{Zijlstra:1993sh}, and massive-quark corrections for $g_1$ are
omitted. SIDIS coefficient functions are implemented up to
NLO~\cite{Furmanski:1981cw, deFlorian:1997zj}, while NNLO corrections are
included via the approximate threshold resummation
formalism~\cite{Abele:2021nyo}. The recent exact polarised coefficient functions
for SIDIS computed in Refs.~\cite{Bonino:2024wgg, Goyal:2024tmo} are not yet
implemented in MAPPDFpol1.0 nor in Ref.~\cite{Borsa:2024mss}. Intrinsic
heavy-quark distributions are considered to be identically zero below their
respective thresholds. Above these thresholds, heavy-quark distributions are
perturbatively generated by means of DGLAP evolution in the zero-mass
variable-flavour-number scheme (ZM-VFNS). Perturbative corrections to the
splitting functions in the DGLAP equations are consistently included up to NNLO
accuracy~\cite{Moch:2014sna, Moch:2015usa,Blumlein:2021ryt}.

At NNLO, the resulting up- and down-quark polarised distributions are well
constrained by data, although sea-quark and gluon distributions remain largely
uncertain and compatible with zero. Despite the inclusion of SIDIS data and NNLO
corrections, the polarised strange asymmetry remains compatible with zero. NNLO
corrections lead to a slight deterioration of the global fit quality, especially
with SIDIS data. Overall, the impact of NNLO corrections on distributions is
moderate, as also confirmed in Ref.~\cite{Borsa:2024mss}. Theoretical
constraints on SU(2) and SU(3) flavour symmetries have minimal effects, but the
positivity constraint significantly affects the shape and uncertainty of PDFs in
the large-$x$ region.

This analysis can be improved on different fronts, starting from the
implementation of the exact SIDIS computations. Indeed, although the terms
common to the threshold-expanded and the complete computation are in excellent
agreement between them, the gluon-gluon channels that open at NNLO in the exact
calculations may affect the polarised gluon. Furthermore, NNLO calculations
introduce significant corrections to the approximate computation in the region
$x \lesssim 0.1$, which may be one of the possible causes for the deterioration
of the fit quality when moving to NNLO. Indeed, the same deterioration is not
observed in Ref.~\cite{Borsa:2024mss}, where a cut $x > 0.12$ is applied to
SIDIS data. Thus, the implementation of the exact calculations is left for
future work.

\section{The current status of polarised PDFs at NLO}
\label{sec:comparison}

Although NNLO accuracy is going to be the new standard for future determinations
of polarised PDFs, the majority of the current available information remains at
NLO. Despite the limited accuracy of NLO analyses, a comparative study of
polarised PDF sets at this level can provide valuable insights resulting from
the choices made in the data sets and in the methodological approaches.
Furthermore, NLO determinations implement exact DIS and, where present, SIDIS
calculations for the structure functions, thereby minimising the dependence on
potential errors associated with approximate perturbative calculations.

Some of the most recent and most widely used sets of polarised PDFs at NLO
accuracy are {\sc MAPPDFpol1.0} ~\cite{Bertone:2024taw},
NNPDFpol1.1~\cite{Nocera:2014gqa}, JAM17~\cite{Ethier:2017zbq}, and
DSSV14~\cite{DeFlorian:2019xxt}\footnote{The list should also include BDSSV22,
the NLO polarised PDF set presented in Ref.~ \cite{Borsa:2024mss}. However,
together with the NNLO determination BDSSV24, these sets are not publicly
available.}. In the following, I provide a brief description for each set listed
above.

\begin{itemize}
  
  \item {\bf NNPDFpol1.1} This set includes open-charm production in
        fixed-target DIS, and jet and W production in proton-proton collisions
        at RHIC, in addition to inclusive DIS data. Target mass corrections are
        also included for the structure function $g_1$. The positivity
        constraint is applied to polarised PDFs, and sum rules are used to
        impose SU(2) and SU(3) flavour symmetries. PDFs are parametrised with a
        neural network and uncertainties are estimated using the Monte Carlo
        replica method.
  
  \item {\bf DSSV14} In addition to inclusive and semi-inclusive DIS data, this
        set takes into account data for jet and hadron production in
        proton-proton collisions at RHIC. A fixed functional form is used to
        parametrise the PDFs and, in the Monte Carlo variant of
        DSSV14~\cite{DeFlorian:2019xxt}, uncertainties are estimated using Monte
        Carlo sampling. The positivity constraint is also enforced to
        distributions, and two parameters are used to account for deviations
        from the assumption of SU(2) and SU(3) flavour symmetries.
  
  \item {\bf JAM17} This set includes data for inclusive and semi-inclusive DIS,
        together with single-inclusive $e^+e^-$ annihilation (SIA) data. This
        set is obtained by performing a simultaneous fit of both PDFs and
        fragmentation functions, both parametrised in terms of a fixed function
        normalised to Euler beta functions. This set does not include positivity
        nor assumptions about SU(2) and SU(3) flavour symmetries. Also in this
        case, uncertainties have been estimated by means of Monte Carlo
        sampling.
\end{itemize}

\begin{figure}[t!]
  \centering
  \includegraphics[width=0.4\textwidth]{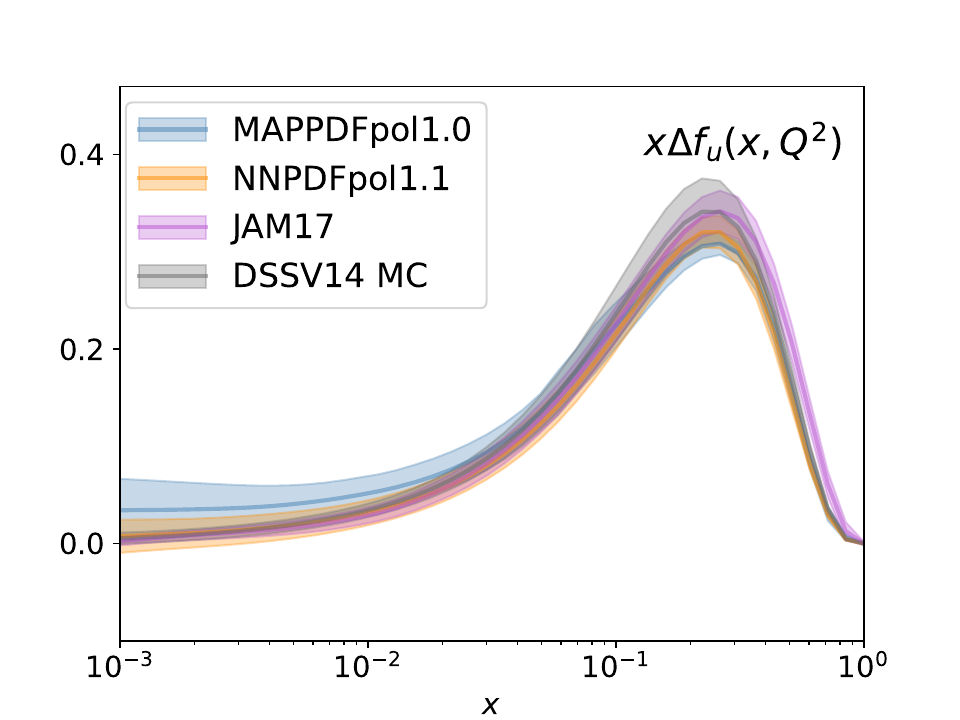}
  \includegraphics[width=0.4\textwidth]{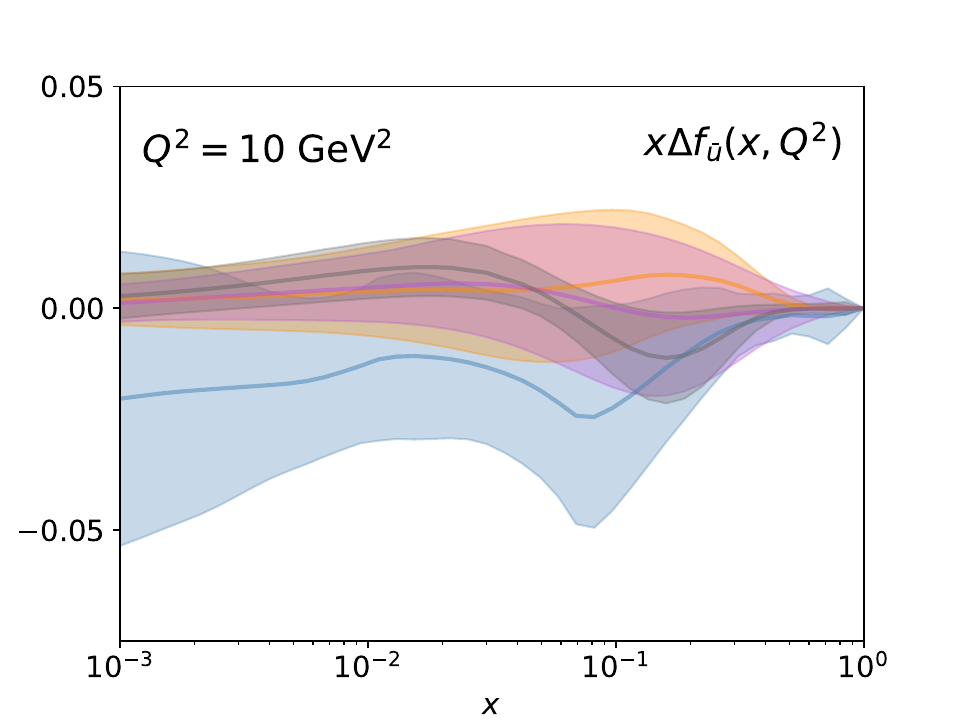}\\
  \includegraphics[width=0.4\textwidth]{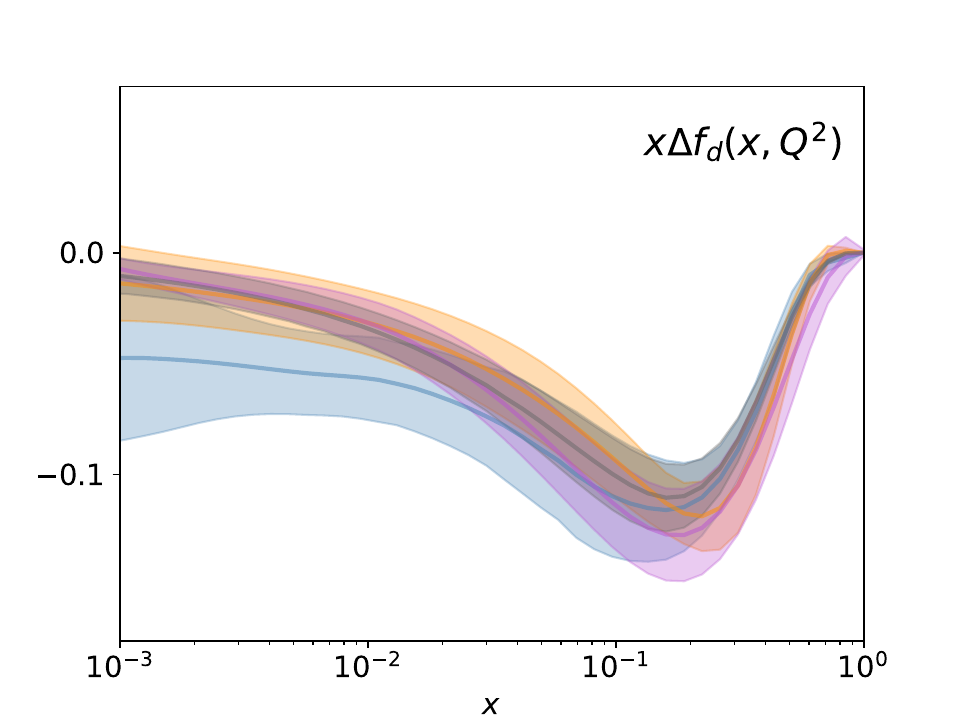}
  \includegraphics[width=0.4\textwidth]{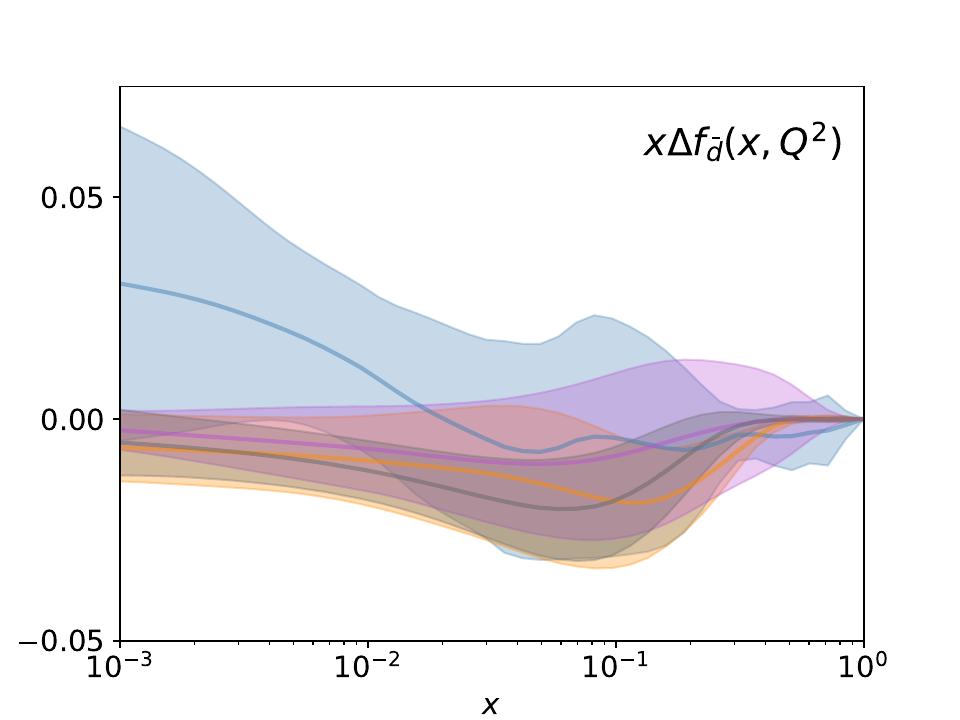}\\
  \includegraphics[width=0.4\textwidth]{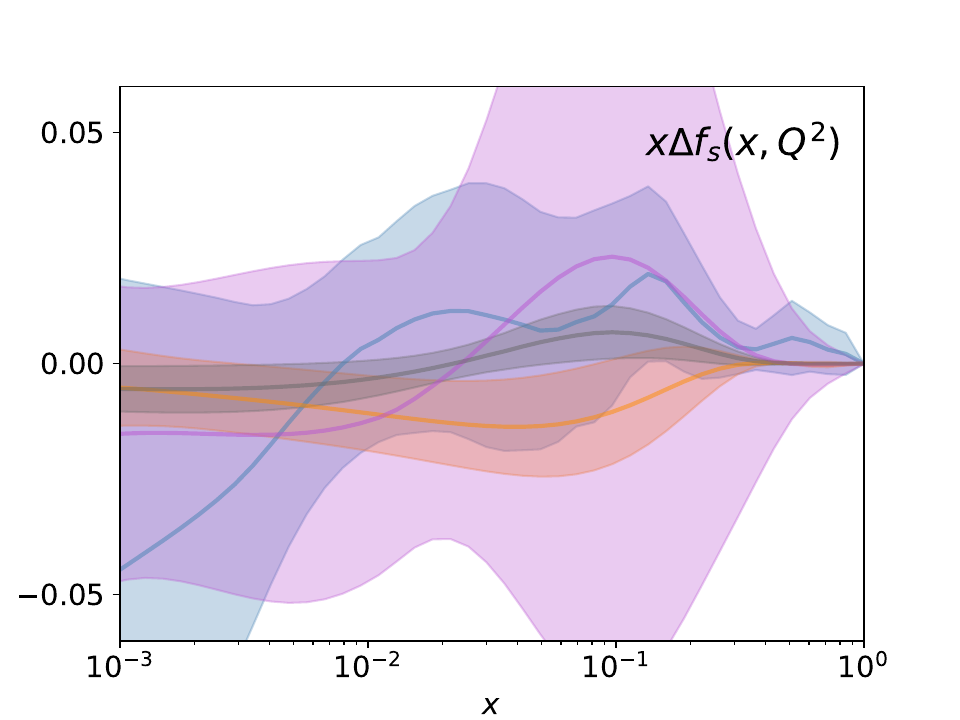}
  \includegraphics[width=0.4\textwidth]{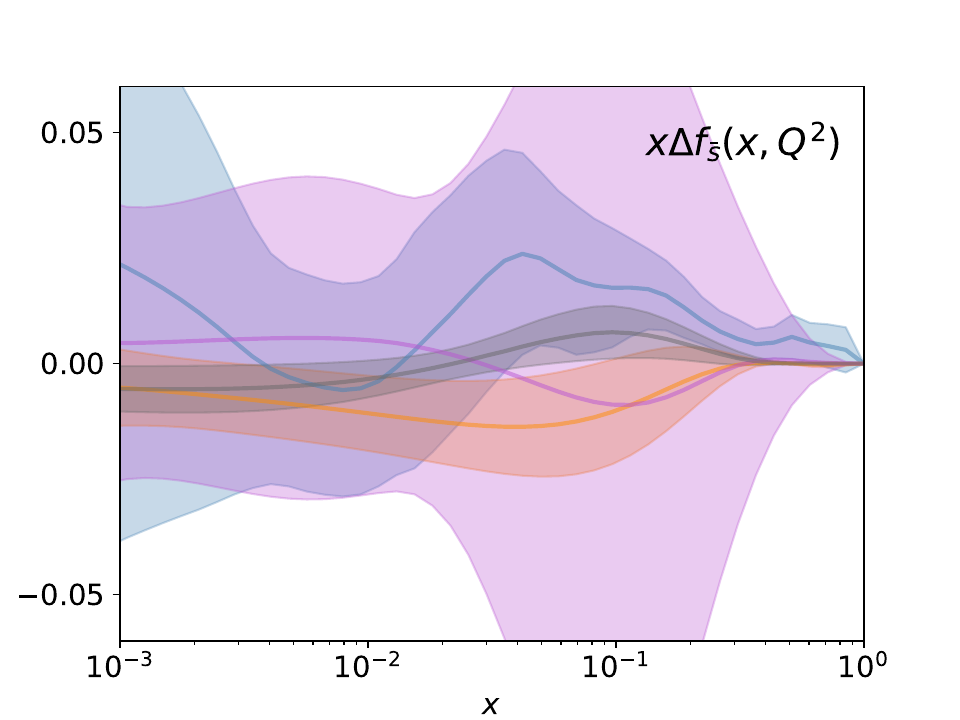}\\
  \includegraphics[width=0.4\textwidth]{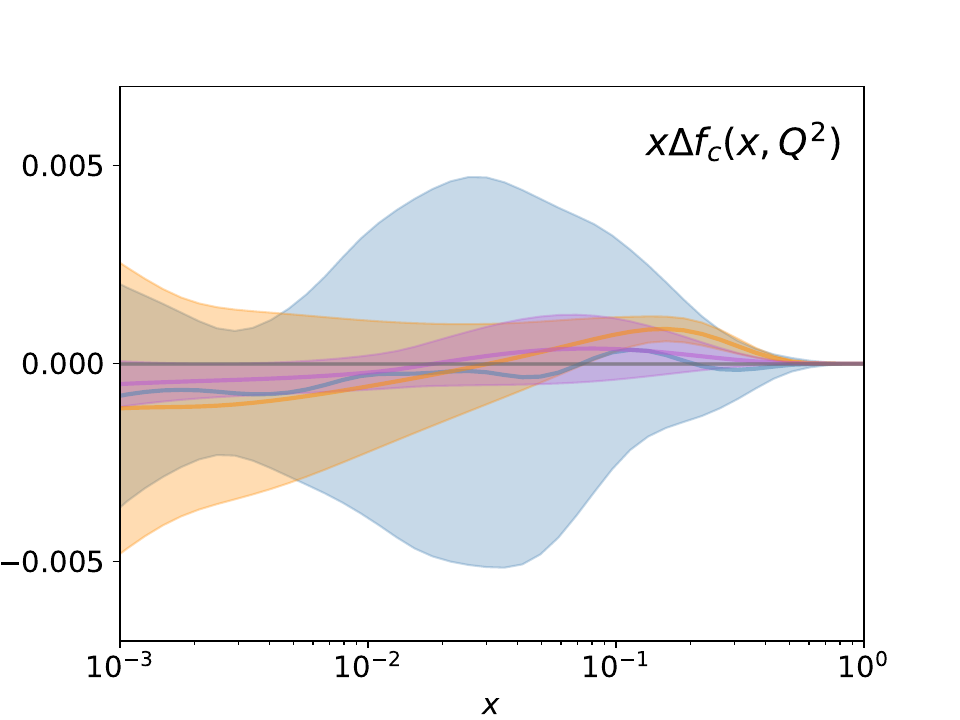}
  \includegraphics[width=0.4\textwidth]{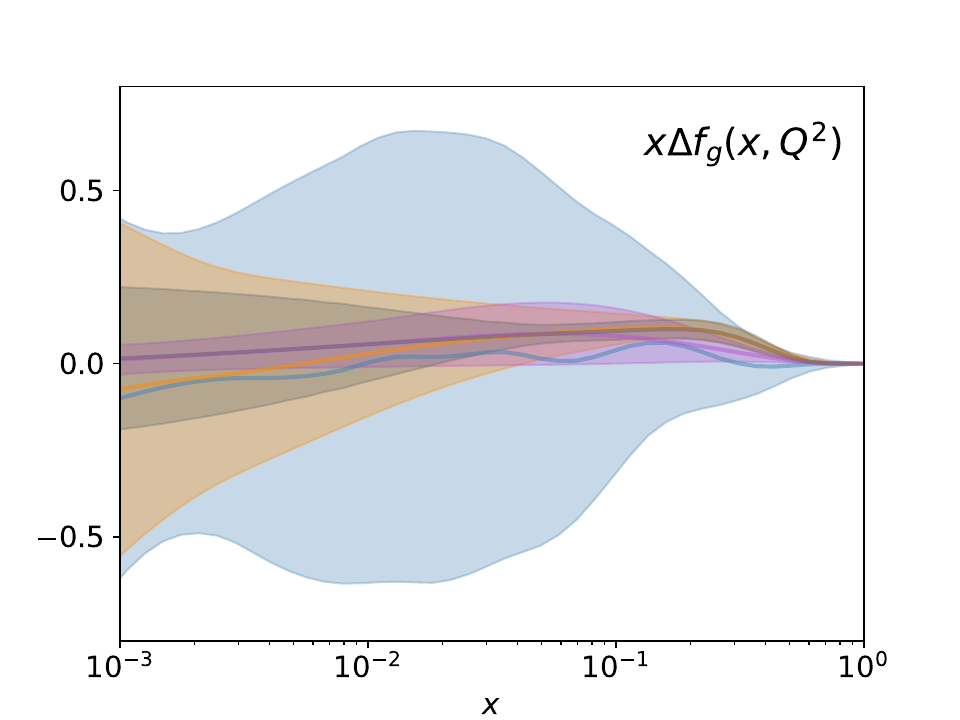}\\
  \vspace{0.5cm}
  \caption{Comparison of the {\sc MAPPDFpol1.0},
  NNPDFpol1.1~\cite{Nocera:2014gqa}, JAM17~\cite{Ethier:2017zbq}, and
  DSSV14~\cite{DeFlorian:2019xxt} polarised PDF sets at NLO. Distributions are
  shown as functions of $x$ at $Q^2=10$~GeV$^2$. Error bands correspond to
  one-sigma uncertainties.}
  \label{fig:MAP_vs_competitors_nlo}
\end{figure}

The aforementioned sets are compared in \cref{fig:MAP_vs_competitors_nlo}, which
shows $\Delta f_u$, $\Delta f_{\bar{u}}$, $\Delta f_d$, $\Delta f_{\bar{d}}$,
$\Delta f_s$, $\Delta f_{\bar{s}}$, $\Delta f_c$, and $\Delta f_g$ as functions of $x$
at the scale $Q^2 = 10$ GeV$^2$. For all PDF sets, error bands correspond to the
one-sigma uncertainty.

The inclusion of gluon-sensitive data helps pin down the polarised gluon PDF, as
it can be seen for NNPDFpol1.1 and DSSV14. RHIC $pp$ data has significant
effects in the region $ x \gtrsim 10^{-3}$, leaving the range $10^{-3} \lesssim
x \lesssim 10^{-2}$ less constrained~\cite{DeFlorian:2019xxt}. These features
are qualitatively consistent between NNPDFpol1.1 and DSSV14. Open-charm
production data, included in NNPDFpol1.1, has been shown~\cite{Nocera:2014gqa}
to have negligible effects on the polarised gluon PDF. The absence of
gluon-initiated data in {\sc MAPPDFpol1.0} is noticeable in the mid-$x$ region
(covered by RHIC data), where uncertainties are larger. On the other hand, JAM17,
albeit having an array of data similar to {\sc MAPPDFpol1.0}, has smaller
uncertainty bands, which may suggest a potential bias in the parametrisation.
Nevertheless, all distributions are compatible within uncertainties. 

The up- and down-quark distributions are generally in good agreement amongst the
different PDF sets. In the low-$x$ region, where DIS and SIDIS data is sparse,
{\sc MAPPDFpol1.0} provides more conservative uncertainties. Deviations are more
noticeable in the case of up- and down-antiquark distributions. In the region $x
\gtrsim 5 \times 10^{-1}$, the $\Delta f_{\bar{u}}$ PDF from {\sc MAPPDFpol1.0}
is pulled closer to DSSV14, which also includes SIDIS data, and away from
NNPDFpol1.1, which does not. This feature was already noted in
Ref.~\cite{Nocera:2014gqa}, where it was observed that the STAR W asymmetry
pulled $\Delta f_{\bar{u}}$ in a different direction than the SIDIS data used in
{\sc MAPPDFpol1.0} and DSSV14.

In the case of the strange PDFs, the compared sets present different behaviours.
Note that only {\sc MAPPDFpol1.0} allows $\Delta f_s$ to be different from
$\Delta f_{\bar{s}}$, and JAM17 parametrises $\Delta f_{\bar{s}}$ and $\Delta
f_s^{+}$. DSSV14 sets $\Delta f_s=\Delta f_{\bar{s}}$, while NNPDFpol1.1 fits
$\Delta f^{+}_s$. In {\sc MAPPDFpol1.0}, JAM17, and DSSV14 the $\Delta f^{+}_s$
PDF is mainly driven by SIDIS data, while the NNPDFpol1.1 constraining power
derives from the inclusion of the semi-leptonic $\beta$-decay parameters. The
effect of the different data sets can be seen in the central-$x$ region, where
deviations between NNPDFpol1.1 and DSSV14 are not contained in the relative
uncertainties. {\sc MAPPDFpol1.0} delivers more conservative uncertainties, and
the shape of $\Delta f_s$ and $\Delta f^{+}_{\bar{s}}$ is more consistent with
that given by the DSSV group. The PDFs from JAM17, on the other hand, are those
that provide the least conservative uncertainties despite using a similar
experimental information as {\sc MAPPDFpol1.0}.

Finally, intrinsic polarised $\Delta c$ is not parametrised in any of the
compared sets. In {\sc MAPPDFpol1.0}, NNPDFpol1.1, and JAM17, polarised $\Delta
c$ is generated by gluon splitting in the perturbative evolution, and indeed
they show similar behaviour as $\Delta g$. Note that the polarised charm PDF in
the DSSV14 Monte Carlo grid has been set to zero.

\section{Conclusions}

The comparative analysis conducted at NLO shows that experimental input and
methodological approach significantly influence the resulting polarised PDFs.
Conversely, the analyses of {\sc MAPPDFpol1.0} and BDSSV24 show that the
differences between the respective NLO and NNLO sets are moderate and smaller
than those observed at fixed perturbative order in
\cref{fig:MAP_vs_competitors_nlo}. This suggests that a similar trend to that
shown in \cref{fig:MAP_vs_competitors_nlo} is likely to be observed across
multiple polarised PDF sets when an analogous comparative analysis will be
undertaken at NNLO level. Therefore, benchmarking across different PDF sets
would be essential to identify the potential sources of the observed
differences, whether they stem from the methodological or experimental input.

\bibliographystyle{JHEP}
\bibliography{DIS2024Proc}
\end{document}